# Abrupt Onset of Second Energy Gap at Superconducting Transition of Underdoped Bi2212

W. S. Lee, I. M. Vishik[1], K. Tanaka[1,2], D. H. Lu[1], T. Sasagawa[1], N. Nagaosa[3], T. P. Devereaux[4], Z. Hussain[2], & Z. -X. Shen[1]

[1]*Department of Physics, Applied Physics, and Stanford Synchrotron Radiation Laboratory Stanford University, Stanford, CA 94305, USA*

[2]*Advanced Light Source, Lawrence Berkeley National Lab, Berkeley, CA 94720, USA*

[3]*Department of Applied Physics, University of Tokyo, Bunkyo-ku, Tokyo 113-8656, Japan*

[4]*Department of Physics, University of Waterloo, Ontario N2L3G1, Canada*

**The superconducting gap–an energy scale tied to the superconducting phenomena-opens on the Fermi surface at the superconducting transition temperature ($T_C$) in conventional BCS superconductors. Quite differently, in underdoped high-$T_C$ superconducting cuprates, a pseudogap, whose relation to the superconducting gap remains a mystery, develops well above $T_C$[1,2]. Whether the pseudogap is a distinct phenomenon or the incoherent continuation of the superconducting gap above $T_C$ is one of the central questions in high-$T_C$ research[3-8]. While some experimental evidence suggests they are distinct[9-18], this issue is still under intense debate. A crucial piece of evidence to firmly establish this two-gap picture is still missing: a direct and unambiguous observation of a single-particle gap tied to the superconducting transition as function of temperature. Here we report the discovery of such an energy gap in underdoped $Bi_2Sr_2CaCu_2O_{8+\delta}$ in the momentum space region overlooked in**

**previous measurements. Near the diagonal of Cu-O bond direction (nodal direction), we found a gap which opens at $T_C$ and exhibits a canonical (BCS-like) temperature dependence accompanied by the appearance of the so-called Bogoliubov quasiparticles, a classical signature of superconductivity. This is in sharp contrast to the pseudogap near the Cu-O bond direction (antinodal region) measured in earlier experiments[19-21]. The emerging two-gap phenomenon points to a picture of richer quantum configurations in high temperature superconductors.**

In spectra taken by angle-resolved photoemission spectroscopy (ARPES), the temperature dependence of the gap appears to be very different between the nodal region and antinodal region, as demonstrated by the Fermi-function divided spectra of UD92K sample shown in Fig. 1(a). Above $T_C$, there is a gapless Fermi arc near the nodal region (C1 to C3 at 120 K); away from this Fermi arc region, the well-known pseudogap gradually takes over and reaches its maximum at the antinodal region (C5 to C8 at 120 K)[1,2,19-22]. Below $T_C$, the magnitude of the pseudogap at the antinode (C8 at 82 K and 10 K) does not show clear temperature dependence across $T_C$, although a sharper peak in the spectrum develops in the superconducting state, as illustrated in Fig. 1(c). The lack of temperature dependence of the antinodal gap size is well known and has been taken as the evidence for pairing above $T_C$[3]. On the other hand, along the Fermi surface (FS) near the nodal region (C1 to C3), an energy gap opens up right below $T_C$, and becomes larger as the system cools to a temperature well below $T_C$. We note that at 82K, there is appreciable thermal population above $E_F$, such that the upper branch of the Bogoliubov dispersion can be clearly seen in the raw spectra for C1-C4, as illustrated in Fig. 1(b). This observation demonstrates that the near nodal gap is related to superconductivity, as Bogoliubov quasiparticles exist only in the coherent superconducting state.

The temperature dependence of the gap evolution near the nodal region of the UD92K sample is analyzed and demonstrated in Fig. 2, suggesting that $T_C$ is correlated with the opening of a single particle gap. Starting with the common procedure of using the symmetrized energy distribution curves (EDCs) at the Fermi crossing point, the data in Fig. 2(c) suggests a collapse of the superconducting gap very close to $T_C$. Shown in Fig. 2(d), the extracted gap size at the locations "A", "B", and "C" is obtained by fitting the symmetrized EDCs to a phenomenological model[23] containing a minimal set of parameters: the gap size and the lifetime broadening of the quasi-particles (see Supplementary Information). As can be seen, the gap at these three locations gradually closes as temperature approaches $T_C$, and vanishes at a temperature close to $T_C$, following the functional form of $\Delta$(T) in weak-coupling BCS theory surprisingly well. The second indication of gap closing comes from the Bogoliubov quasiparticle dispersion as shown in Fig. 2(a) and (b). The shift of the Bogoliubov peak suggests that the superconducting gap size rapidly reduces when the temperature approaches $T_C$. Further, the disappearance of the Bogoliubov band within a narrow temperature range (87-97 K) having small thermal broadening difference confirms the collapse of the gap above $T_C$ (see Supplementary Information). Thirdly, we remark that this sudden onset of the gap is closely related to the abrupt drop of the quasiparticle lifetime at $T_C$, as shown in the fitted $\Gamma$ from our data (Fig. S2(b) in Supplementary Information). This observation is consistent with microwave spectroscopy[24], thermal Hall conductivity[25], and ARPES data of nodal quasiparticles[26,27]. Taken together, the temperature dependence of the electronic states in the nodal region is consistent with a BCS superconductor.

We note that such behavior has not been observed in previous ARPES measurements[23] because this nodal "BCS-gap" region has been overlooked due to insufficient momentum space sampling. The gap at previously reported momentum position has already opened above $T_C$ due to the proximity to the pseudogap region, whose behavior was extrapolated to the entire Fermi surface[20,23], leading to the conclusion of pseudogap and superconducting gap being the same gap. Our finding indicates that such extrapolation and its conclusion should be revised.

To study the temperature evolution of the gap function, spectra at three selected temperatures were recorded to accommodate a large number of cuts along the Fermi surface. As shown in Fig. 3(a), the gap along the Fermi surface can be roughly divided into two groups. One group is the region near the node (C1-C2) where the gap is temperature dependent with an onset temperature very close to $T_C$. The other group is associated with the antinodal region (C6-C7), which does not show any significant temperature dependence across $T_C$. As we move toward the antinode (including curves C3-C5), the temperature dependence of the gap across $T_C$ becomes less pronounced, implying a smooth transition from one group to the other. With these two rather different temperature variations, an nontrivial temperature dependent evolution of the gap function $|\Delta_k(T)|$ along the Fermi surface can be sketched. As shown by the 82 K data in Fig. 3(b), a gap consistent with a simple $d_{x^2-y^2}$ form, $|\cos k_x - \cos k_y|/2$, begins to develop near the node at a temperature right below $T_C$, whereas the gap near the antinode deviates from this nodal region *d*-wave gap. Surprisingly, when the system is cooled well below $T_C$, the momentum dependence of the gap along the entire FS appears to be consistent with the simple $d_{x^2-y^2}$ form, at least for this doping. This non-trivial temperature evolution is another surprise associated with the discovery of the superconducting gap near the nodal region. We note that the values of $2\Delta_{k=antinode}(T = 10\ K) / k_B T_C$ of this *d-wave* gap is approximately 9, which is still much larger than the value ~ 4.12 predicted by weak-coupling *d*-wave BCS theory.

In Fig. 3(c) and (d), data of an underdoped sample with $T_C$ = 75 K and overdoped sample with $T_C$ = 86 K are shown. The temperature dependence of the gap function is consistent with the UD92K sample except that the gapless region above $T_C$ extends with increasing doping. This change of the gap function in the superconducting state is qualitatively different from the simple mean field behavior with a temperature independent pairing interaction, where the momentum dependence of the gap should not change at temperatures below $T_C$. Thus, the observed temperature dependent evolution of the gap function implies an intriguing relation between the superconducting gap and pseudogap. We note that in heavily underdoped samples, where the pseudogap is much more pronounced than the superconducting gap, the gap function can only evolve into a “U” shape at our lowest achievable temperature (see Fig. 3(c) and also Ref. 17). In Fig. 4, we summarize schematically the temperature dependent evolution of the gap function in the three samples with different doping level we have studied.

It seems impossible to explain our data within a single gap picture. We are not aware of any mechanism that would create an energy gap which opens at different temperatures on the same sheet of the Fermi surface. In addition, the temperature dependent evolution of the gap function along FS in the superconducting state also seems very difficult to reconcile within a single gap picture. It appears more reasonable to assume the existence of two energy gaps. The energy gap opening at $T_C$ near the nodal region is associated with the order parameter of the superconducting state, while the pseuodogap near the antinodal region represents an energy scale associated with a different mechanism that may or may not be related to superconductivity. This is consistent with the two-gap picture derived from the doping dependence measurements on heavily underdoped Bi2212 described in our recent work[17]. Notably, a number of other spectroscopy experiments, such as Andreev reflection[11], intrinsic tunnelling spectroscopy[13], and femtosecond spectroscopy[14], were also interpreted as a gap opening at $T_C$, which was difficult to understand in the context of previous ARPES[19-21] and STM[22] results dominated by the antinodal region (see Supplementary Information). We argue that these experimental probes are more sensitive to the nodal region, where the gap opens at $T_C$, as demonstrated in our data. As a remark, the temperature dependence of the STM spectrum has been re-visited recently and the coexistence of two energy gaps in the underdoped cuprates has been suggested[15,16]. In particular, a normalization procedure enables the revelation that one of the gaps disappears at $T_C$[16], consistent with our finding near the nodal region.

What is the relationship between the observed two gaps? On the one hand, the distinct temperature and doping dependence seem to suggest a competing nature between the nodal BCS-like superconducting gap and the antinodal pseudogap; on the other hand, the evolution of the gap profile into a simple *d*-wave form at low temperature for UD92K and OD86K samples seems to suggest an intimate relationship between the superconducting gap and the pseudogap. Theoretical calculations, in which the pseudogap is ascribed to a charge density wave competing with the superconducting state[6,7], demonstrate a similar temperature dependence and doping dependence of the gap profile shown in this paper and our recent study on heavily underdoped system[17]. Likewise, theories, which treat pseudogap as preformed Cooper pairs, could also predict a two-gap behavior. For example, a bipolaron theory[28] has demonstrated two energy gaps with distinct temperature dependence: one gap opens at $T_C$ in a BCS fashion and the other is essentially temperature independent across $T_C$. However, these theories would require the two gaps to add in quadrature, something seems inconsistent with the antinodal data at least for UD92K and OD86K sample. The other theories[3-5] have yet developed detailed temperature dependence for a direct comparison with this experiment. In all cases, our finding would put a strong constraint on theory.

**Supplementary Information** accompanies the paper on www.nature.com/nature.

**Acknowledgements** We thank Rob Moore for experimental assistance and D. J. Scalapino, S. Kivelson and T. K. Lee for helpful discussions. ARPES experiments were performed at the Stanford Synchrotron Radiation Laboratory (SSRL) which is operated by the Department of Energy Office of Basic Energy Science.

**Author Information** The authors declare no competing financial interests. Correspondence and request for materials should be addressed to Z.X. Shen (zxshen@stanford.edu) or Wei-Sheng Lee (leews@stanford.edu).

**Figure 1 Temperature and momentum dependence of the low energy excitations in slightly underdoped Bi2212 ($T_C$ = 92 K).** (a) Image plots of the Fermi-Dirac function divided ARPES spectrum (see Supplementary Information) along Fermi surface taken at three different temperatures: above $T_C$ (102 K), right below $T_C$ (82 K), and well below $T_C$ (10 K). The high intensity region represents the band dispersion along the cutting directions as indicted by solid white lines in panel (d). In the FD-divided spectrum, this high intensity region either breaks at the Fermi energy ($E_F$) if there is an energy gap, or passes $E_F$ if there is no detectable gap. (b) Raw energy distribution curves (EDCs) near the Fermi crossing point ($k_F$) for C1-C4 at 82 K. The short vertical lines indicate the thermally-populated Bogoliubov band above $E_F$. The Bogoliubov band dispersion is a signature of the superconducting state. (c) Temperature dependence of raw EDCs near $k_F$ of C8. Short vertical lines indicate the gap energy. Momentum position of EDCs is indicated by the dashed lines in the panel C8 of (a). A sharp peak in the spectrum can be observed right below $T_C$, although the gap size remains about the same across Tc. (d) A partial FS mapping measured at 102 K in a quadrant of the first Brillouin zone.  Map is obtained by integrating raw spectra over the energy window $E_F$ ±10 meV and symmetrizing with respect to the diagonal of the Brillouin zone. The intensity near the nodal direction (white dashed line) is suppressed because of the matrix element effect under this experimental setup.

**Figure 2 Detailed temperature dependence of the superconducting gap near the nodal region of underdoped Bi2212 ($T_C$ = 92 K) measured under two different experimental configurations.** Data shown in upper panels of (a)-(d) were measured along cuts parallel to the (π,0)-(π,π) direction, using 22.7 eV photons and an energy resolution of 5 meV. The data in the lower panels were

measured along cuts parallel to the (0,0)-($\pi$,$\pi$) direction, using 7 eV photons and an energy resolution of 3.2 meV. (a) Temperature dependence of raw EDCs near the FS crossing points “A” and “C”. The short bars indicate the peak of the thermally-populated upper Bogoliubov band, whose position at 70 K is marked by the dashed line as a reference. When the temperature approaches $T_C$, the peak of the thermally-populated Bogoliubov band moves toward $E_F$, and then disappears above $T_C$. (b) Temperature dependence of the peak position of the thermally-populated upper Bogoliubov band in the raw spectra shown in (a), which extrapolates to zero near $T_C$. The behaviors demonstrated in (a) and (b) suggest that the magnitude of the superconducting gap decreases when the temperature approaches $T_C$, and vanishes above $T_C$. (c) Temperature dependence of the symmetrized EDCs at FS crossing points “A” and “C” superimposed on their fit(black curves) to the phenomenological model[23]. At temperatures above $T_C$, the two peaks of the symmetrized EDCs merge into one peak, suggesting the collapse of the gap. (d) Temperature dependence of the fitted gap size at FS crossing points near the node, as indicated in the insets. The dashed lines show the temperature dependence of the superconducting gap based on weak-coupling BCS theory and serve as a guide-to-the-eye for our data. The $\Delta_k(T = 0)$ for the BCS curves are adjusted independently in order to fit the data at different locations. The error bar is estimated to be ± 2 meV accounting for the uncertainties from the fitting procedure (± 0.5 meV), the Fermi energy calibration (± 0.5 meV), and additional 100 % margin.

**Figure 3 Temperature dependence of the gap profile** (a) The fitted gap values from the UD92K sample for selected Fermi surface crossings as defined in Fig. 1(d). Note the discrete temperature points should not be regarded as a gap

closing at 102 K. (b) The fitted gap values versus the simple $d_{x^2-y^2}$ functional form, $|\cos k_x - \cos k_y|/2$. The lines are guides-to-the-eyes indicating the expected momentum dependence of a simple $d_{x^2-y^2}$ form. (c) and (d) show the temperature dependent evolution of the gap function for an underdoped sample with $T_C$ = 75 K and an overdoped sample with $T_C$ = 86 K, respectively. The error bars for the superconducting gap are the same as those described in the caption of Fig. 2, while those of pseudogap are larger because of the larger uncertainty from the fitting procedure.

**Figure 4 Schematic illustrations of the gap function evolution for three different doping levels.** At 10 K above $T_C$, there exists a gapless Fermi arc region near the node; while a pseudogap has already fully developed near the antinodal region (red curve). With increasing doping, this gapless Fermi arc elongates (thick red curve on FS), as the pseudogap effect weakens. At $T<T_C$, a *d*-wave like superconducting gap begins to open near the nodal region (green curve), however, the gap profile in the antinodal region deviates from the simple $d_{x^2-y^2}$ form. At a temperature well below $T_C$ ($T<<T_C$), the superconducting gap with the simple $d_{x^2-y^2}$ form eventually extends across entire Fermi surface (blue curve) for the UD92K and OD86K samples, but not for the UD75K system.

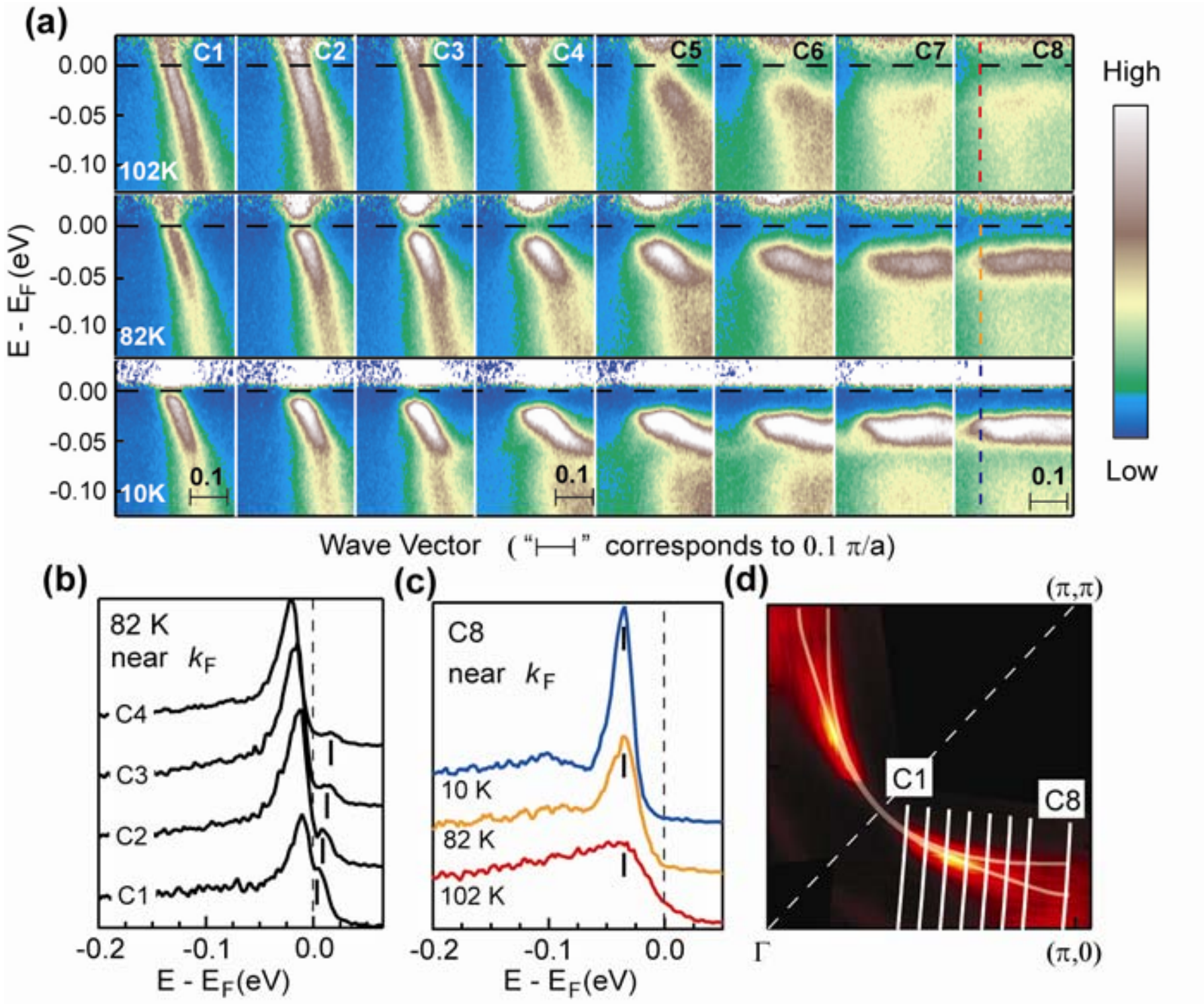

(a)
C1
C2
C3
C4
C5
C6
C7
C8
102K
82K
10K
E - E_F (eV)
0.00
-0.05
-0.10
0.1
High
Low
Wave Vector ( "⊢⊣" corresponds to 0.1 π/a)
(b)
82 K
near k_F
C4
C3
C2
C1
-0.2 -0.1 0.0
E - E_F(eV)
(c)
C8
near k_F
10 K
82 K
102 K
-0.2 -0.1 0.0
E - E_F(eV)
(d)
(π,π)
C1
C8
Γ
(π,0)

Figure 1

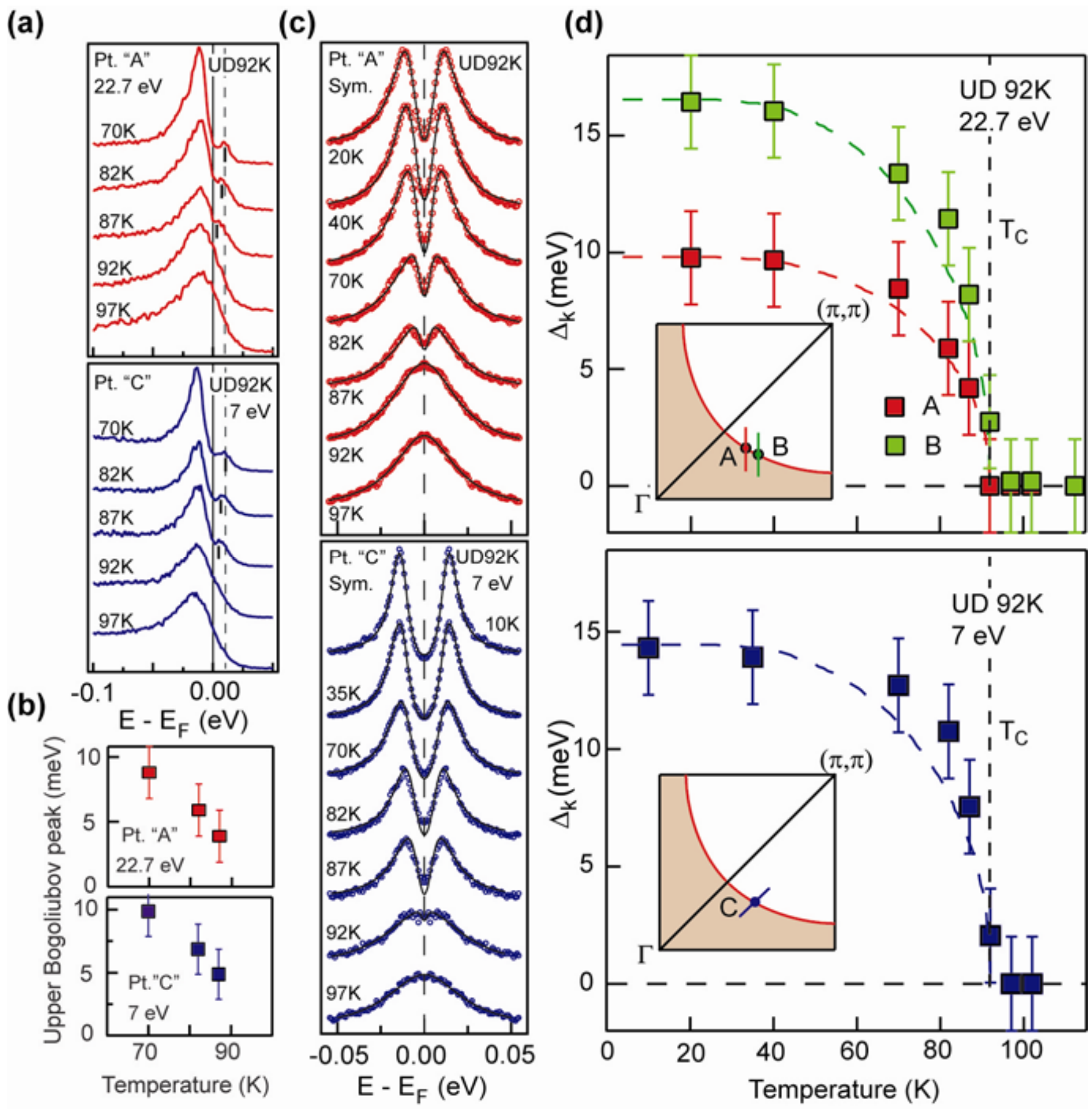

(a)
Pt. "A"
22.7 eV
UD92K
70K
82K
87K
92K
97K
Pt. "C"
UD92K
7 eV
E - E_F (eV)
(b)
Upper Bogoliubov peak (meV)
Pt. "A"
22.7 eV
Pt."C"
7 eV
Temperature (K)
(c)
Pt. "A"
Sym.
UD92K
20K
40K
10K
35K
E - E_F (eV)
(d)
UD 92K
22.7 eV
T_C
Δ_k (meV)
(π,π)
Γ
A
B
UD 92K
7 eV
C
Temperature (K)

Figure 2

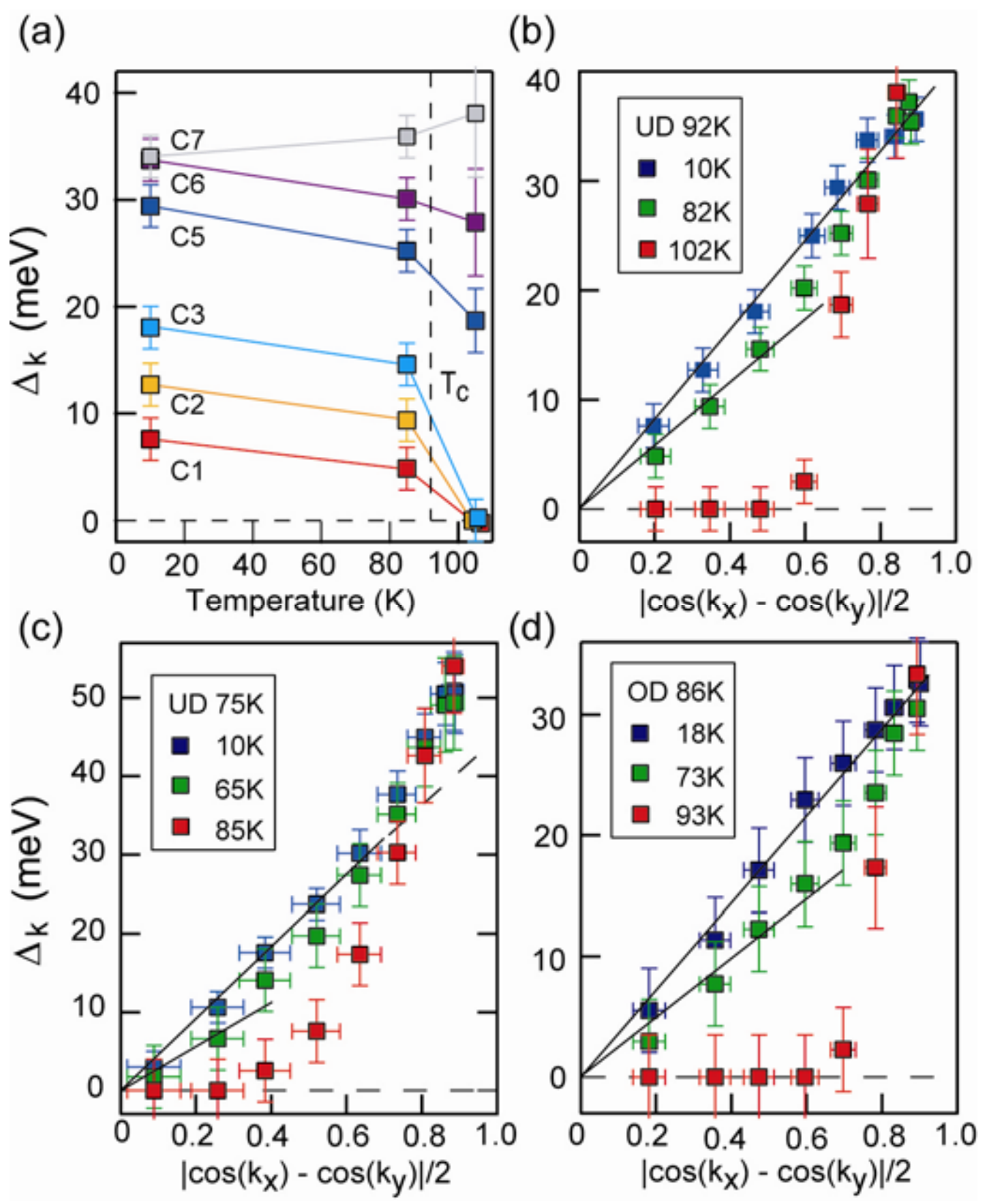
(a)
(b)
(c)
(d)
C7
C6
C5
C3
C2
C1
T_c
Δk (meV)
Temperature (K)
|cos(kx) - cos(ky)|/2
UD 92K
10K
82K
102K
UD 75K
10K
65K
85K
OD 86K
18K
73K
93K

Figure 3

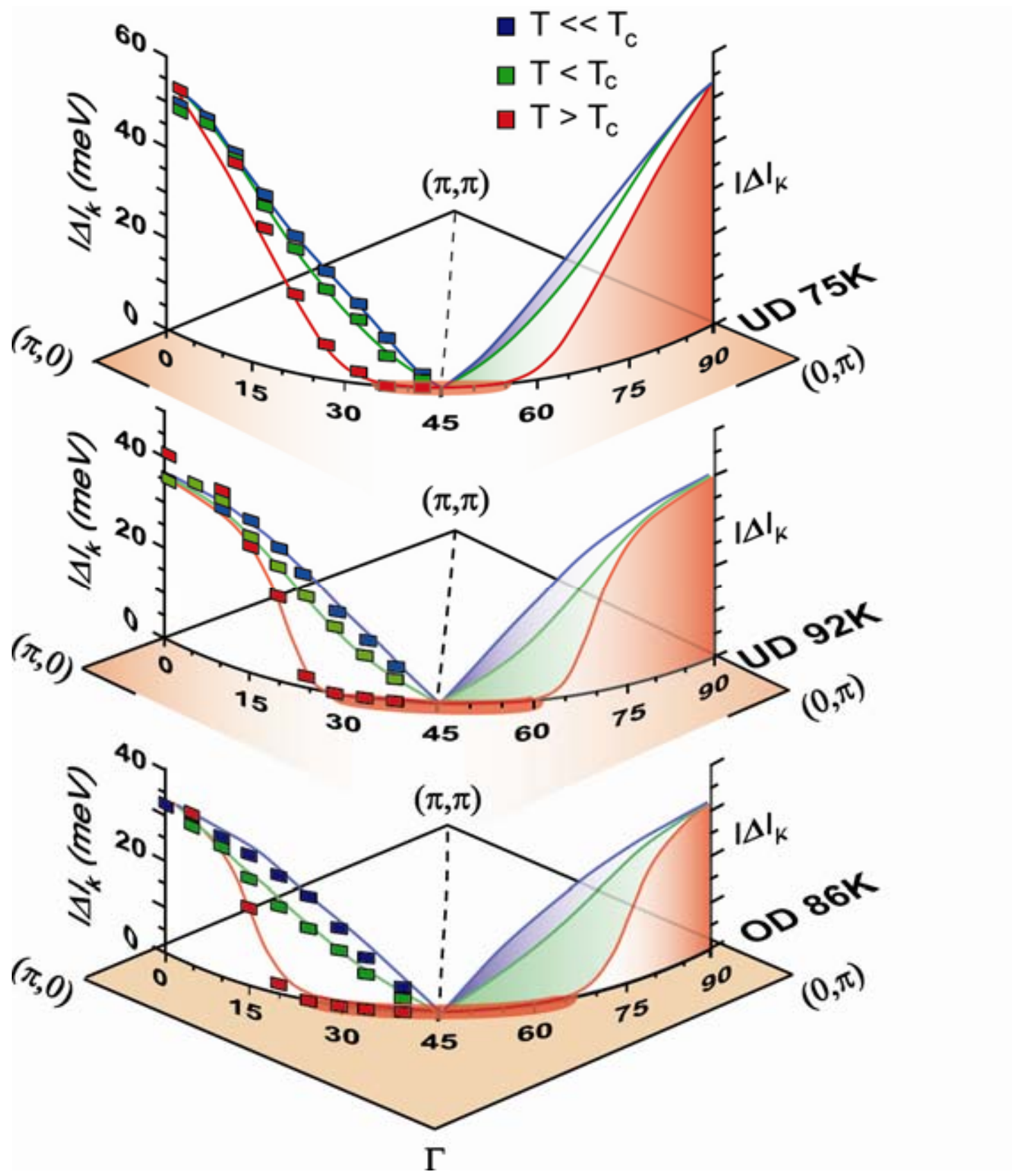

Figure 4